\documentstyle[aps]{revtex}
\begin{document}
\title{A first-order transition with power-law singularity in models with 
absorbing states\\}
\author{Adam  Lipowski \cite{byline}\\}
\address{Department of Physics, A. Mickiewicz University, Ul. Umultowska 85, 
61-614 Pozna\'{n}, Poland}
\date{\today}
\maketitle
\begin{abstract}
We study one- and two-dimensional models which undergo a transition 
between active and absorbing phases.
The transition point in these models is of novel type: jump of the order 
parameter coincides with its power-law singularity.
Some arguments supported  by Monte Carlo simulations prompted us to predict
the exact location of the transition point.
Both models possess gauge-like symmetry.
\end{abstract}
\pacs{05.70.Ln}
\section{Introduction}
Recently, nonequilibrium phase transitions have been intensively studied in 
variety of models~\cite{HAYE1}.  
In addition to some potential applications, the motivation to study
these transitions comes from the belief that they can be categorized into 
universality classes similarly to equilibrium phase transitions. 
In this context, models which exhibit transitions between active and 
absorbing phases~\cite{DICK} are of particular interest.  
There already exists substantial numerical evidence that phase transitions 
in such models indeed can be classified into some universality
classes.  
For example, it is believed that models with unique absorbing states
should belong to the so-called directed-percolation (DP) universality
class~\cite{GRASSJAN}.  
Moreover, models with double (symmetric) absorbing
states or with some conservation law in their dynamics, belong to
another universality class~\cite{PARITY}.

Similarly to equilibrium systems, nonequilibrium continuous phase transitions 
are not the only possibility - some models are known to undergo discontinuous
transitions~\cite{ZIFF,ALBANO,EVANS}.  
Although such transitions are not classified into universality classes, they 
might be more relevant since, discontinuous transitions are, at present, the 
only type of transitions which can be observed experimentally. 
On the contrary, the experimental realization of continuous phase transitions 
still remains an open problem~\cite{HAYE}.

One reason for a relatively good understanding of equilibrium phase
transitions is a wealth of exactly solvable models in this field~\cite{BAXTER}.
With this respect, the situation is much worse for nonequilibrium phase
transitions.  
None of the models with absorbing states and with continuous or discontinuous 
transitions was solved exactly and all results concerning 
the critical exponents or the location of a transition point are only 
numerical. 

In the present paper we study certain models with
infinitely many absorbing states.
At a certain value of a control parameter $r=r_{{\rm c}}$, these models 
undergo a transition between active and absorbing phases. 
But the interesting point is a novel type of this transition: it seems to 
combine some features of both discontinuous and continuous transitions.
Namely, at $r=r_{{\rm c}}$ an order parameter jumps discontinuously to 
zero but in addition to that the order parameter has a power-law singularity 
upon approaching the transition point from the active phase.
Moreover, some elementary arguments, supported by Monte Carlo simulations,
prompted us to predict the exact location of the transition point in both 
models, namely, $r_{{\rm c}}=0$.
Both models have a gauge-like symmetry, which might be responsible for the
unusual behaviour of these models.
\section{Square lattice}
Our first model is a certain modification of a model introduced in a context 
of modelling biological evolution~\cite{LIPLOP,LIP99}.
It is defined on a two-dimensional ($d=2$) square lattice where for each bond 
between the nearest-neighbouring sites $i$ and $j$ we introduce bond variables 
$w_{i,j}\in (-0.5,0.5)$.
Introducing the parameter $r$, we call the site $i$ active when 
$\prod_{j} w_{{i,j}}<r$, where $j$ runs over all nearest neighbours of $i$.
Otherwise, the site is called nonactive.
The model is driven by random sequential dynamics and when the active site 
$i$ is selected, we assign anew, with uniform probability, four bond 
variables $w_{i,j}$, where $j$ is one of the nearest neighbours of $i$.
Nonactive sites are not updated, but updating a certain (active) site might
change the status of its neighbours.

An important quantity characterizing this model is the steady-state density 
of active sites $\rho$.
How does $\rho$ change with the control parameter $r$?
Of course, for $r\geq (0.5)^4=0.0625$ all sites are active ($\rho=1$) for any 
distribution of bond variables $w_{i,j}$.
It is natural to expect that for $r<0.0625$ and not too small there will be a 
certain fraction of active sites ($\rho>0$) and this fraction will decrease 
when $r$ decreases.
Since the dynamical rules imply that the model has absorbing states with all 
sites nonactive ($\rho=0$), one can expect that at a 
certain $r$ the model undergoes a transition between the active and absorbing 
phases.
On general grounds one expects that this transition might be either 
continuous and presumably of (2+1)DP universality class~\cite{COM1} or 
discontinuous.

The existence of a transition is confirmed in Fig.\ref{f1}, which shows the 
density $\rho$ as a function of $r$ obtained using Monte Carlo simulations.
The simulations were performed for the linear system size $L=300$ and we
checked that the presented results are, within small statistical error, 
size-independent.
After relaxing the random initial configuration for $t_{{\rm rel}}=10^4$, we 
made measurements during runs of $t=10^5$ (the unit of time is defined as a 
single on average update/lattice site).
From this figure one can also see that the transition point $r_{{\rm c}}$ 
is located very close to $r=0$ and in the following we are going to show that 
it is very likely that in this model $r_{{\rm c}}=0$ (exactly).

First, we show that for $r<0$ the model is in the absorbing phase.
The argument for that is elementary and based on the following observation:
for $r<0$ there exists a finite probability that after updating a given site 
will become nonactive forever.
Indeed, when one of the anew selected bonds satisfies the condition
\begin{equation}
|w_{i,j}|<-r/(0.5)^3,
\label{e1}
\end{equation}
then the sites $i$ and $j$ become permanently nonactive (i.e., no matter what 
are the other bonds attached to these sites, they will always remain 
nonactive).
For $r<0$ there is a finite probability to satisfy Eq.~\ref{e1} and the 
above  mechanism leads to the rapid decrease of active sites and hence 
the system reaches an absorbing state.
The above mechanism is not effective for $r\geq 0$ since there is no value 
which would ensure permanent nonactivity of a certain site.

To confirm that for $r<0$ the system is in the absorbing phase, we present in 
Fig.~\ref{f2} the time evolution of $\rho$ for $r=-10^{-6}$ and $-10^{-7}$.
Although these values are very close to $r=0$, one can clearly see that the 
system evolves toward the absorbing state.
(For $r$ smaller than these values, the approach to the absorbing state would 
be even faster.)
As we have already mentioned, for $r\geq 0$ the mechanism which generates 
permanently
nonactive sites is not effective.
Most likely, this has important consequences: as shown in Fig.~\ref{f2}, even 
for $r=0$ the system does not evolve toward the absorbing state but remains in 
the active phase.

These results indicate that at $r=0$ the model undergoes a first-order 
transition between active and absorbing phases, characterized by a 
discontinuity of the order parameter $\rho$.
However, the most interesting feature of the model is the fact that upon 
approaching the first-order transition point $r=0$ the order parameter 
exhibits a power-law singularity.
Such singularities usually signals a continuous transition.
This singularity, which is already visible in the inset of Fig.~\ref{f1}, is 
also presented in the logarithmic plot in Fig.~\ref{f3}.
The parameter $\rho_0=0.359$ (i.e., the density of active sites for $r=0$) in 
Fig.~\ref{f3} was obtained from the least-square analysis of small-$r$ 
($r\leq 10^{-3}$) data shown in Fig.~\ref{f1} using the formula
\begin{equation}
\rho(r)=\rho_0+Ar^{\beta},
\label{e2}
\end{equation}
where we assumed that the critical point is located at $r=0$~\cite{COM2}.
From the slope of the data in Fig~\ref{f3}, we estimate $\beta=0.58(1)$, 
which might  suggest that the exponent $\beta$ for that model is the same as 
in the (2+1)DP~\cite{VOIGT,COM2}.
However, a characteristic feature of models of the DP universality class 
is that at the transition point the model falls into an absorbing state.
Our model at the transition point ($r=0$) is not in the absorbing phase 
(see Fig.~\ref{f2}), but it enters the absorbing phase as soon as $r$ 
becomes negative.
In addition, scaling behaviour of our numerical data persists on a relatively 
small interval of $r$ and asymptotically a different behaviour might sets in.
Further arguments against DP criticality of this model are presented in the 
next section.

Let us notice that the above model is characterized
by very large gauge-like symmetry.  
Indeed, inverting ($\pm$) four bond variables around any elementary square 
does not change the activity of sites.
The gauge symmetry was examined for many equilibrium lattice 
models~\cite{KOGUT}.
However, the up-to-now examined models with absorbing states do not possess 
this kind of symmetry.
It would be interesting to check whether the unusual properties of this model 
are related with this symmetry.
In the following we examine a one-dimensional model which possesses a 
similar symmetry.
\section{Triangular ladder}
Let us examine a model defined on a one-dimensional ($d=1$) ladder-like 
lattice, where each site also has four neighbours (see Fig.~\ref{zigzag}).
When defined with the same dynamical rules as the model examined in the 
previous section, this $d=1$ model also has an analogous gauge symmetry 
(see Fig.~\ref{zigzag}).

We examined the properties of this model using the same Monte Carlo method.
Results of our simulations for the steady-state density $\rho$ are shown in 
Fig.~\ref{f1} and Fig.~\ref{f3}.
As is usually the case of models with absorbing states, Monte Carlo 
simulations of the $d=1$ version are much more accurate.
For example, close to and at the transition point $r=0$ we simulated the system
of the size $L=3\cdot 10^5$ and the simulation time was typically $t=10^6$.
As a result we were able to probe a much closer vicinity of the transition 
point.

Our results indicate that the behaviour of $d=1$ and $d=2$ versions of
this model is very similar.
Both models exhibit a qualitatively the same transition at $r=0$.
In the $d=1$ case our estimations of the critical parameters are:
$\rho_0=0.314827$ and $\beta=0.66(3)$.
A relatively good scaling  behaviour in this case is confirmed on over two 
decades (see Fig.~\ref{f3}).
The obtained value of the exponent $\beta$ clearly excludes the DP value (in
the case of one-dimensional DP $\beta=0.276486$~\cite{JENSEN}.

To get further insight into the nature of the transition point we examined 
the size dependence of the relaxation time $\tau$.
We measured the time needed for the system starting from the random
initial configuration to reach the steady-state.
Typically, at the continuous transition $\tau$ diverges as $L^z$ where $z$ is 
a positive exponent.
For the one- and two-dimensional DP $z=1.58$ and 1.76, respectively.
At the discontinuous transition one expects that $\tau$ remains finite in the
thermodynamic limit (i.e., $z=0$).
For $r=0$ and $d=1$ results of our measurements, shown in Fig.~\ref{tau} are, 
in our opinion, inconclusive. 
They may suggest a power-law divergence with a small exponent $z (\sim 0.2$),
but positive curvature of our data might asymptotically lead to $z=0$.
On the other hand even if $z=0$, it is not certain whether $\tau$
remains finite or diverges, but slower than a power of $L$.
For $r>0$ (i.e., off-criticality) the numerical data are similar to the $r=0$,
but on general grounds one expects that $\tau$ remains finite in the 
thermodynamic limit.
\section{Summary}
In the present paper we studied two models which exhibit remarkably similar and
unusual behaviour.
These models have a transition point which, although mainly of discontinuous 
nature (jump of the order parameter and $z=0$) has a certain feature of 
continuous transitions (power-law singularity of the order parameter).

The main weakness of our paper is the lack of any theoretical argument which
would explain the behaviour of these  models.
Both models possess certain gauge-like symmetry.
The role of such symmetry in models with absorbing states was not yet 
explored and it is possible that the behaviour of these models might be 
related with this symmetry.

Are there any indications that such transitions might take place in real 
systems?
In our opinion, one of the possible applications might be related with phase 
transitions in nuclear physics.
Indeed, there are some indications that multifragmentation of heavy nuclei 
resembles a phase transition which has both first- and second-order 
features~\cite{NUCLEI}. 
Such systems have been already modeled using Ising-like models.
However, such an approach implicitly assumes a thermalization of the system, 
which is not obvious in these multifragmentation processes.
Models with absorbing states might provide an alternative description 
of such processes.
\acknowledgements
I thank prof.~Des Johnston for his hospitality and the Department of 
Mathematics of the Heriot-Watt University (Edinburgh, Scotland) for allocating 
computer time.
I also thank H.~Hinrichsen for interesting discussion.

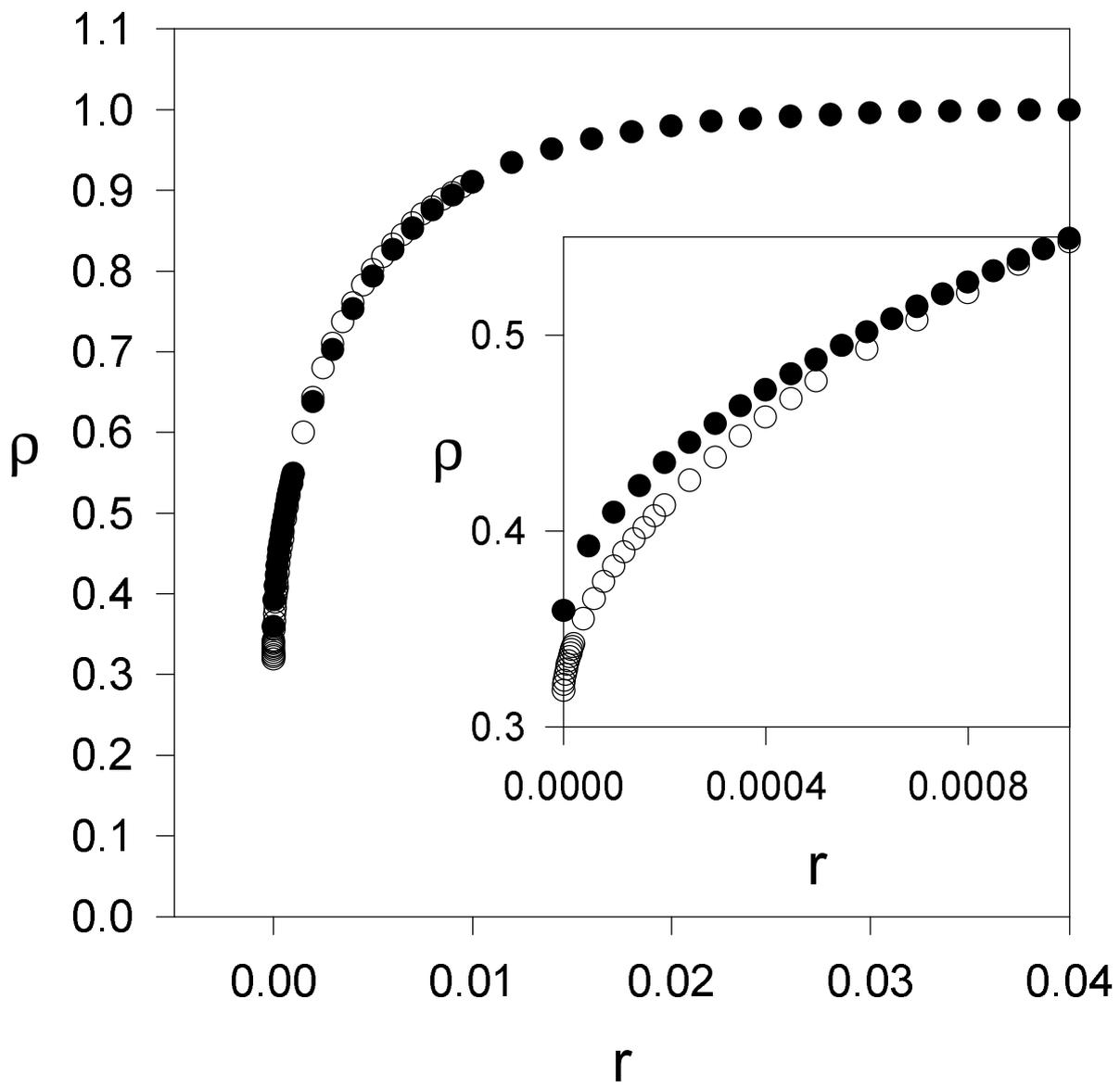
\begin{figure}
\caption{The steady-state density of active sites $\rho$ as a function of $r$
for the $d=1$ ($\circ$) and $d=2$ ($\bullet$) models.
The simulations were made for the system size $L=300$ ($d=2$) 
and up to $L=3\cdot 10^5$ ($d=1$).
For $r<0$ the system quickly reaches an absorbing state (see 
Fig.~\protect\ref{f2}).
The error bars are smaller than the plotted symbols.
The inset shows our data in the vicinity of $r=0$.
For $r>10^{-3}$ the density $\rho$ for both models is almost the same.}
\label{f1}
\end{figure}
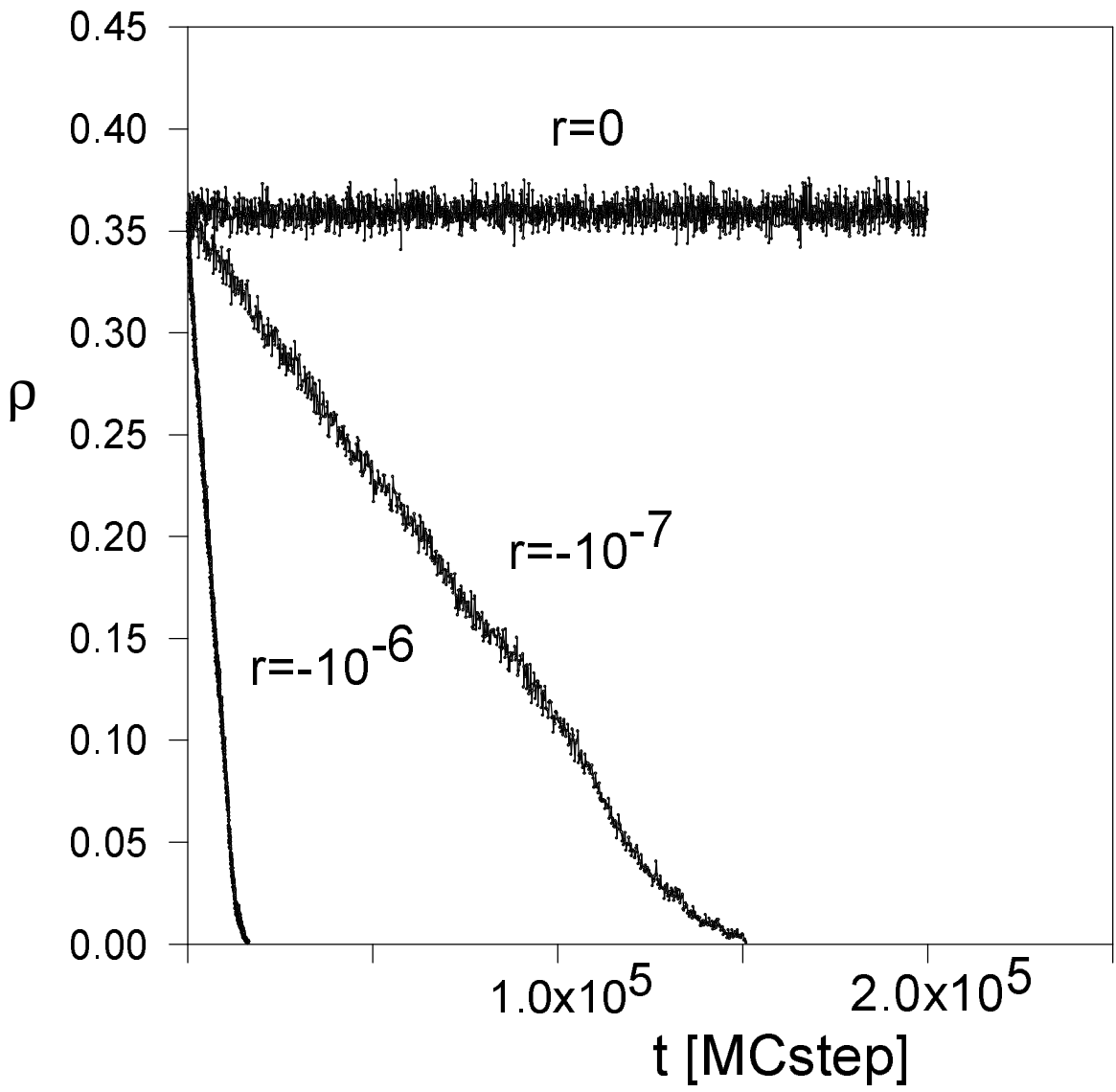
\begin{figure}
\caption{The time evolution of the density of active sites $\rho(t)$ ($d=2$ 
and $L=100$).
For $r=0$ the system remains in the active phase, but as soon as $r$ 
becomes negative it evolves toward an absorbing state.
For $r=0$ and $L=300$ the simulations give, within small statistical errors, 
the same results.}
\label{f2}
\end{figure}
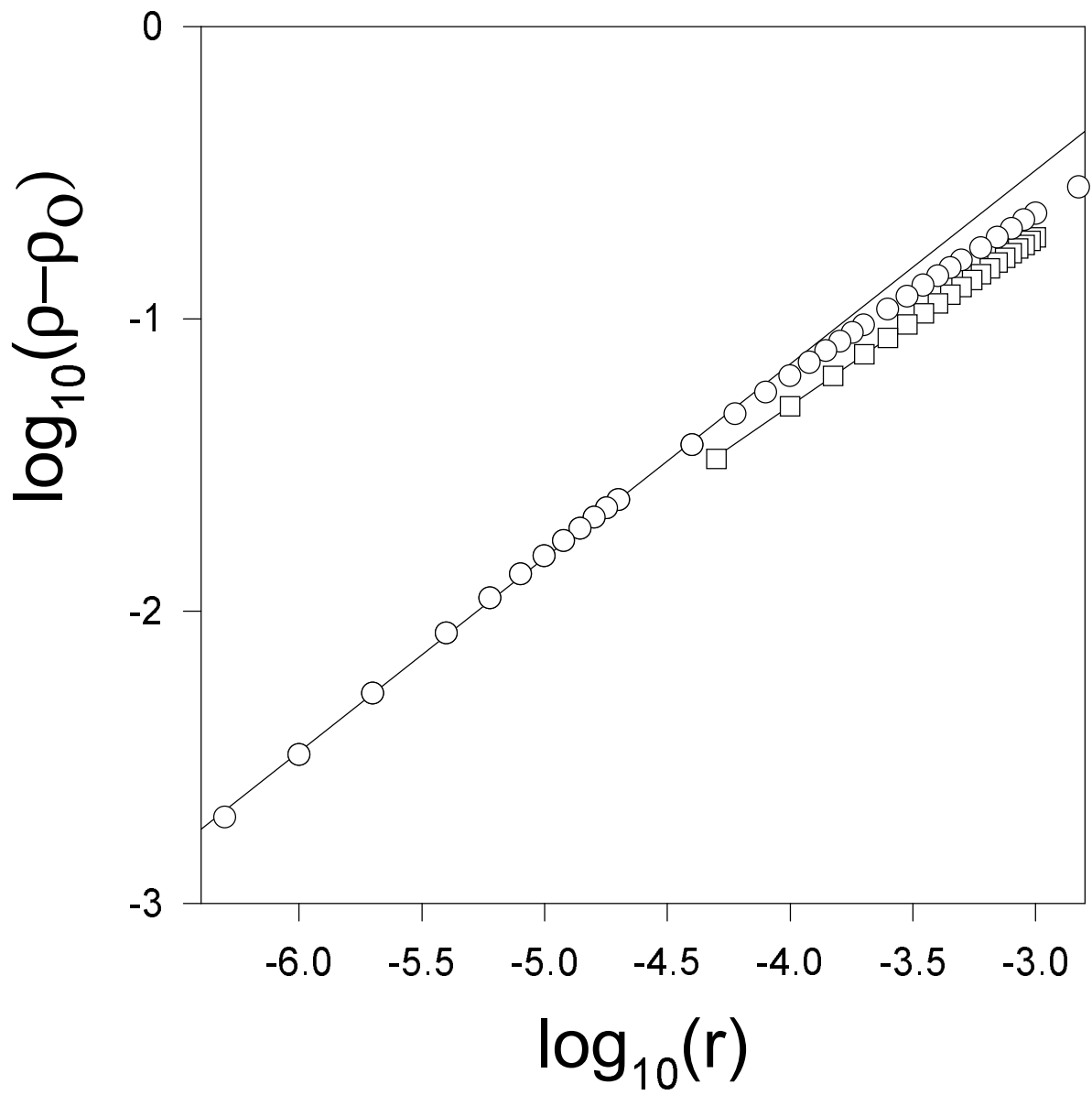
\begin{figure}
\caption{The plot of log$_{10}(\rho-\rho_0)$ as a function of log$_{10}(r)$ 
with $p_0=0.359$ ($d=2$, $\Box$) and $p_0=0.314827$ ($d=1$, $\circ$).
The lines have slopes corresponding to $\beta=0.58$ ($d=2$) and $\beta=0.66$ 
($d=1$).}
\label{f3}
\end{figure}
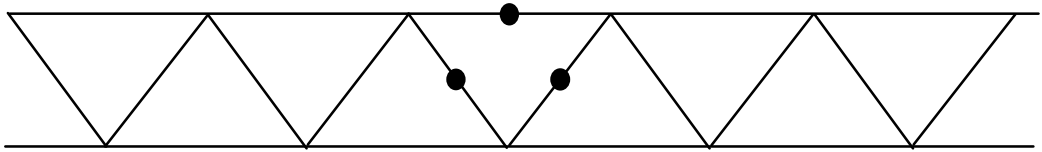
\begin{figure}
\caption{The triangular ladder.
When three bonds ($\bullet$) around a certain triangle are 
inverted, activity of the system remains unchanged.}
\label{zigzag}
\end{figure}
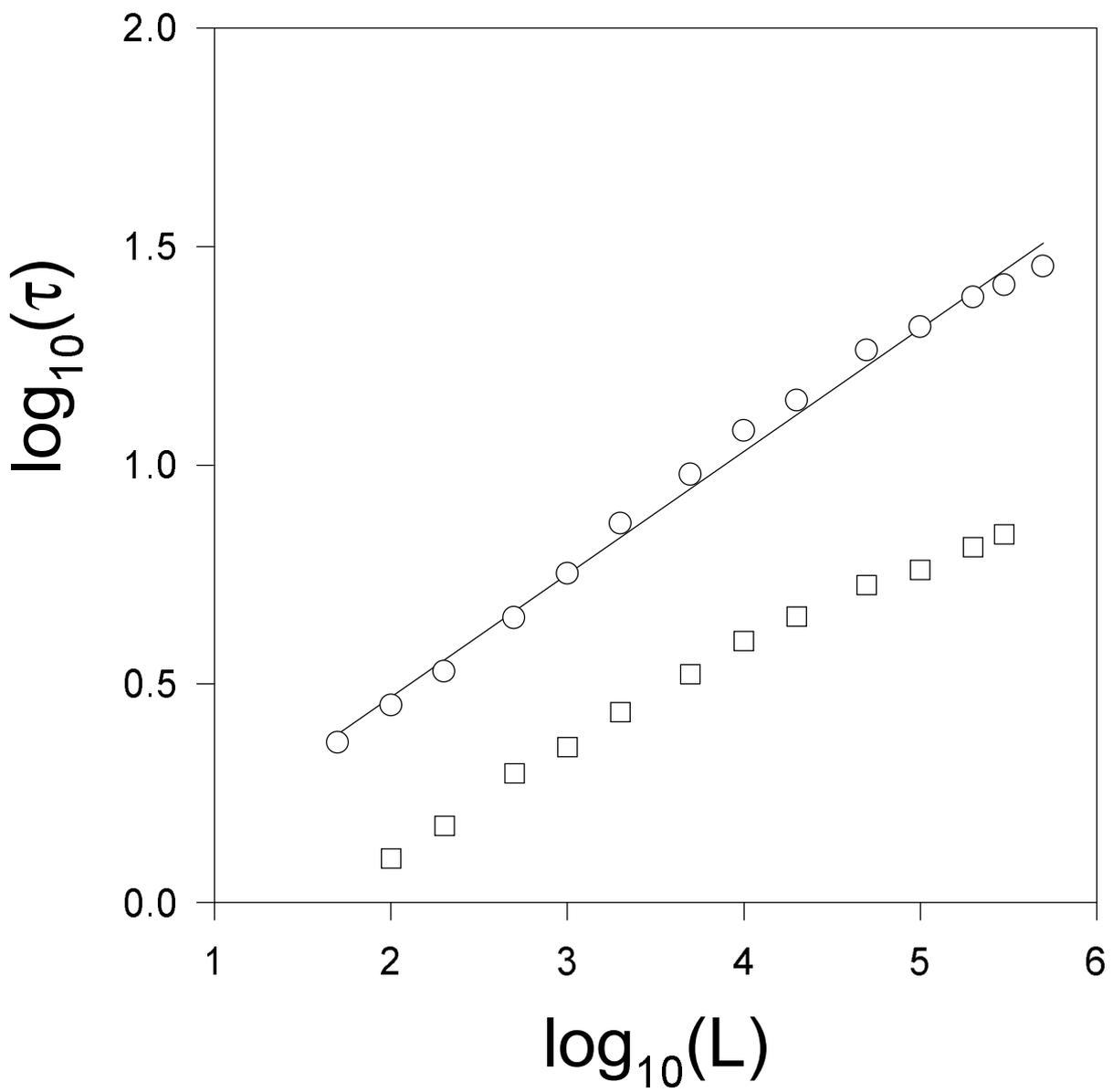
\begin{figure}
\caption{The size dependence of the relaxation time $\tau$ for the 
$d=1$ model and $r=0$ ($\circ$) and $10^{-3}$ ($\Box$).
The straight line has a slope 0.3.}
\label{tau}
\end{figure}

\begin{references}
\bibitem{COM1} Recently, it has been shown that some two-dimensional models 
with a continuous transition and infinitely many absorbing states 
do not belong to the (2+1)DP universality class~\cite{LIP99}.
\bibitem{COM2} We also fitted the data shown in the inset of Fig.~\ref{f1} 
using the formula $\rho(r)=\rho_0+A(r-r_{{\rm c}})^{\beta}$ with unspecified 
$\rho_0, A, r_{{\rm c}}$ and $\beta$.
We obtained $r_{{\rm c}}\sim 10^{-6}, \rho_0\sim 0.358$ and $\beta\sim 0.55$.
\end{references}
\end{document}